\documentclass[aps,prx,superscriptaddress,twocolumn,10pt]{revtex4-2}
\usepackage[utf8]{inputenc}
\usepackage[titletoc,toc,title,page]{appendix}

\usepackage{csquotes}
\usepackage[italian,english]{babel}

\usepackage{microtype} 
\usepackage{bm} % provides more bold symbols

\usepackage{dsfont} % provides more \mathbb \smallskipymbols
\usepackage{amsmath,amssymb,amsthm,thmtools}
\usepackage{mathtools}
\usepackage{cases}
\usepackage{calc}
\usepackage{mathrsfs} % provides the nice \matchsrc font
\usepackage[normalem]{ulem} %provides \sout command for striking through text

\newcommand\redsout{\bgroup\markoverwith{\textcolor{red}{\rule[0.5ex]{2pt}{1.pt}}}\ULon}
\usepackage{parskip}
\usepackage{nameref}
\usepackage[colorlinks=true]{hyperref}
\usepackage[nameinlink]{cleveref}
\crefname{appsec}{Appendix}{Appendices}
\crefname{box}{Box}{Box}
\hypersetup{
  colorlinks   = true, %Colours links instead of ugly boxes
  urlcolor     = green!80!black, %Colour for external hyperlinks
  linkcolor    = blue, %Colour of internal links 	q
  citecolor    = red!80!black %Colour of citations
}

\usepackage{physics} % provides Dirac notation and lots of other things

\usepackage{float} % provides the option H for includegraphics

\usepackage{graphicx}

\usepackage[usenames,dvipsnames,table]{xcolor}
\usepackage{easyReview}

\usepackage{tikz}
\usetikzlibrary{calc,shapes.geometric}

\usepackage{placeins}
\usepackage{multirow,tabularx,booktabs}
\usepackage[most]{tcolorbox}
\newtcbtheorem{tbox}{Box}{enhanced, float*=t, width=\textwidth, label type=box}{box}

\usepackage[printonlyused,withpage,nohyperlinks,smaller]{acronym}

\graphicspath{{./figures/}}

\newcommand{\PP}{\mathbb{P}}

\newcommand{\calH}{\mathcal{H}}

\newcommand{\parTitle}[1]{\textit{#1 ---}}

\newcommand{\addLI}[1]{{\color{blue}#1}}

\renewcommand{\addLI}[1]{#1}

\begin{document}

%\title{An operative definition of quantum information scrambling}
\title{Quantum scrambling via accessible tripartite information}

% \author{Gabriele Lo Monaco}
% \author{Luca Innocenti}
% \author{Dario Cilluffo}
% \author{Diana A. Chisholm}
% \author{Salvatore Lorenzo}
% \author{G. Massimo Palma}

% check affiliation?
\author{Gabriele Lo Monaco}
\let\comma,
%\affiliation{Institut de Physique Th\'eorique\comma{} CEA Saclay\comma{} CNRS UMR 3681\comma{} F-91191 Gif-sur-Yvette\comma{} France}
\affiliation{Universit\`a degli Studi di Palermo\comma{} Dipartimento di Fisica e Chimica - Emilio Segr\`e\comma{} via Archirafi 36\comma{} I-90123 Palermo\comma{} Italy}
\author{Luca Innocenti}
\let\comma,
\affiliation{Universit\`a degli Studi di Palermo\comma{} Dipartimento di Fisica e Chimica - Emilio Segr\`e\comma{} via Archirafi 36\comma{} I-90123 Palermo\comma{} Italy}
\author{Dario Cilluffo}
\affiliation{Institut f\"ur Theoretische Physik and IQST\comma{} Albert-Einstein-Allee 11\comma{} Universit\"at Ulm\comma{} 89069 Ulm\comma{} Germany}
\author{Diana~A.~Chisholm}
\let\comma,
\affiliation{Centre of Quantum Materials and Technologies\comma{} School of Mathematics and Physics\comma{} Queen's University\comma{} Belfast BT7 1NN\comma{} United Kingdom}
\author{Salvatore Lorenzo}
\affiliation{Universit\`a degli Studi di Palermo\comma{} Dipartimento di Fisica e Chimica - Emilio Segr\`e\comma{} via Archirafi 36\comma{} I-90123 Palermo\comma{} Italy}
\author{G. Massimo Palma}
\affiliation{Universit\`a degli Studi di Palermo\comma{} Dipartimento di Fisica e Chimica - Emilio Segr\`e\comma{} via Archirafi 36\comma{} I-90123 Palermo\comma{} Italy}
\affiliation{NEST\comma{} Istituto Nanoscienze-CNR\comma{} Piazza S. Silvestro 12\comma{} 56127 Pisa\comma{} Italy}

\begin{abstract}
    \noindent
    Quantum information scrambling (QIS), from the perspective of quantum information theory, is generally understood as local non-retrievability of information evolved through some dynamical process, and is often quantified via entropic quantities such as the tripartite information.
    We argue that this approach comes with a number of issues, in large part due to its reliance on quantum mutual informations, which do not faithfully quantify correlations directly retrievable via measurements, and in part due to the specific methodology used to compute tripartite informations of the studied dynamics.
    We show that these issues can be overcome by using accessible mutual informations, defining corresponding ``accessible tripartite informations'', and provide explicit examples of dynamics whose scrambling properties are not properly quantified by the standard tripartite information.
    Our results lay the groundwork for a more profound understanding of what QIS represents, and reveal a number of promising, as of yet unexplored, venues for futher research.
\end{abstract}

\maketitle

\section{Introduction}

% \parTitle{What is QIS}
Motivated by the longstanding problem of the black hole information paradox~\cite{Hawking:1976ra},
``Quantum information scrambling'' (QIS)~\cite{Hayden:2007cs} was originally proposed as a way to reconcile the impossibility of retrieving information that fell in a black hole, with the fact that quantum information cannot be destroyed~\cite{braunstein_quantum_2007}.
More generally, a lot of attention has been devoted to understanding what features of a dynamical process hinder the retrievability of information evolving through it ~\cite{hosur2016chaos,swingle2018unscrambling,xu2022scrambling}.
This kind of QIS is present in various many-body systems~\cite{iyoda2018scrambling,pappalardi2018scrambling}, is considered a hallmark feature of quantum chaos and thermalization~\cite{zhou2017operator,Sekino:2008he,lashkari2013towards,banuls2017dynamics},
and may play a role to reach quantum advantage~\cite{Rossini:2019nfu,Gyhm:2021gzx}.

The approaches used to quantify QIS are divided between out-of-time-ordered correlators (OTOCs)~\cite{larkin1969quasiclassical,swingle2016measuring,nakamura2019universal,bergamasco2019out,xu2022scrambling,garcia2021quantum,landsman2019verified,yoshida2019disentangling,ahmadi2022quantifying,hosur2016chaos,swingle2018unscrambling} and entropic methods based on mutual informations~\cite{zhuang2022phase,alba2019quantum,touil2020quantum,touil2021information}, and tripartite information~\cite{yeung1991new,cerf1998information,kitaev2006topological,casini2009remarks,hayden2013holographic,rota2016tripartite,hosur2016chaos,schnaack2019tripartite,iyoda2018scrambling,pappalardi2018scrambling,seshadri2018tripartite}.
We will focus here on the latter.
\addLI{A standard approach is to quantify QIS via the tripartite information of the auxiliary state obtained evolving a maximally entangled state through the target dynamics.
However, this approach does not always faithfully quantify how much the input information is hidden in the correlations between different output subsystems, which is what we aim to capture with QIS measures in the first place~\cite{hosur2016chaos,swingle2018unscrambling,xu2022scrambling}.
More specifically, we recognize several important shortcomings of tripartite-information-based quantifiers.
First, the tripartite information, defined using quantum mutual informations, contains terms originating from the quantum discord~\cite{ollivier2001quantum} which do not correspond to correlations observable via local measurements on the output states.
Furthermore, the tripartite information is not sensible to the choice of encoding and decoding bases, and as we will argue with several explicit examples, it is not possible to accurately capture the idea of information scrambling in quantum systems in a basis-independent way.
}
Based on these considerations, we propose a novel framework to quantify QIS through \textit{accessible} mutual information in order to obtain a clear interpretation of the QIS features of a given dynamics.
% \addRev{
% This can lead to situations where there is a stark contrast between the scrambling as measured by the tripartite information, and the scrambling as measured using accessible quantities. \addG{This discrepancy is in part due to quantum discord contribution affecting the standard tripartite information; in contrast, the proposed accessible tripartite information only quantifies, by construction, the amount of input information encoded into {\it classical} correlations among output subsystems.}
% }
% \cLI{.}
\addLI{This new definition highlights how the scrambling properties of an evolution strongly depend on how information is encoded in the input states, and furthermore solves the discrepancy that was due to quantum discord contributions in the standard tripartite information.}
Within this framework, we find that different output subsystems can be optimally correlated with the input states with respect to different encoding bases.
This immediately raises interesting questions about whether asking how information about input states is encoded locally in different output states is well-posed.
Understanding these underappreciated aspects of information spreading is crucial to deepen our understanding of the phenomenon of QIS. 

% \addG{The basis-independent nature of the standard tripartite information is another source of discrepancy with the proposed new framework.}
% questo l'abbiamo già detto sopra no?

% {\color{ForestGreen}
% We recognize two important shortcomings of the standard tripartite information as a measure of QIS: (1) it is crucial to mind which output subsystem is been considered, and (2) the tripartite information contains terms due to quantum discord corresponding to both input and output encoding bases.
% We find that to talk about scrambling one needs to specify measurement bases and what output systems are been looked at. This highlights the complexity and richness of the phenomenon of quantum information scrambling, which cannot be captured using a single quantifier such as the tripartite information.
% The tripartite information is not sensible to the choice of encoding and decoding bases. As we will show explicitly with several examples, it is not possible to capture the idea of information scrambling in quantum systems in a basis-independent way.

% Chiarire ruolo di discord. Discord base dependent. Explicitly state that the presence of quantum discord is the cause of discrepancies between tripartite and accessible quantities.
% Possibly also that to really access these QMIs one needs entanglement between input and output.
% }

\parTitle{Outline}
We start in~\cref{sec:issues_I3} by reviewing the tripartite information and its applications to QIS, with a focus on the abovementioned issues. We then discuss in~\cref{sec:accessible_I3} a direct way to address them by replacing QMIs with their accessible counterparts.
Finally in~\cref{sec:encoding_bases} we study how the optimal input encoding depends on the choice of measured output subsystem.
% \cLI{a sto punto forse possiamo unirlo con paragrafo sopra, che gia sta facendo un outline}

\section{Measuring QIS via $I_3$}
\label{sec:issues_I3}

% \parTitle{Summary of section}
A common method to quantify the QIS of a unitary evolution $U$ is via the tripartite information of a state $\rho_U$ that captures the relevant features of the dynamics.
However, as we will show via a number of explicit examples, this quantifier may be misleading
due to the discordant~\cite{ollivier2001quantum} terms arising in its definition, which cause it to not always directly quantify accessible correlations, and the definition of $\rho_U$ making the definition of QIS for multipartite output states unclear.

\parTitle{Outline}
We start introducing the definition and main properties of $I_3$ in~\cref{subsec:definition_of_I3}.
Accessibility issues intrinsic to $I_3$ are discussed in~\cref{subsec:accessibility_issues_I3}.
Finally, in~\cref{subsec:single-arm,subsec:double-arm} we disuss specifically the common ways $I_3$ is used to quantify QIS, and highlight some issues intrinsic to these approaches.
%\addDM{(Alte D.: \textit{Are we sure we need a subsummary for each section? Maybe removing this one and merging the following section with the latter could avoid us to repeat the definition of $I_3$.})}
%\cLI{a me le outlines piacciono sinceramente, danno struttura al discorso}

\subsection{Definition of $I_3$}
\label{subsec:definition_of_I3}

% \parTitle{Quantum $I_3$}
Given a tripartite state $\rho$ in a space $\calH_A\otimes\calH_B\otimes\calH_{C}$, its tripartite information is defined as
\begin{equation}\label{eq:definition_tripartite_info}
    % I_3(\rho) \equiv I(\rho_{A,B})
    % + I(\rho_{A,C})
    % - I(\rho_{A,BC}),
    I_3(\rho) \equiv
    I(A:B)_\rho + I(A:C)_\rho - I(A:BC)_\rho,
\end{equation}
where $I(A:B)_\rho$ is the QMI between $A$ and $B$,
\begin{equation}
    I(A:B)_\rho \equiv S(\rho_A)+S(\rho_B)-S(\rho_{AB}),
\end{equation}
and $S(\cdot)$ is the von Neumann entropy.
An equivalent expression for $I_3$ is
\begin{equation}
\begin{gathered}
    I_3(\rho) = S(\rho_A) + S(\rho_B) + S(\rho_C) \\
    - S(\rho_{AB}) - S(\rho_{AC}) - S(\rho_{BC})
    + S(\rho_{ABC}),
\end{gathered}
\end{equation}
which highlights its symmetry with respect to the three subspaces.

Another feature of $I_3$ is that for any pure four-partite state $\ket{\psi}\in\calH_{ABCD}$, all the tripartite informations computed on reduced tripartite states are identical \cite{rota2016tripartite}, that is,
\begin{equation}
    I_3(\rho_{ABC})
    = I_3(\rho_{ABD})
    = I_3(\rho_{ACD})
    = I_3(\rho_{BCD}).
\end{equation}

One possible intuition underlying $I_3$ is that it quantifies the correlations between $A$ and $BC$ which cannot be accessed via local measurements on $B$ and $C$.
In other words, it quantifies the amount of information about $A$ that is ``hidden'' in the correlations between $B$ and $C$.
If $I_3<0$, then $BC$ might be expected to contain more information about $A$ than $B$ and $C$ do, if taken separately.
It was further argued~\cite{hosur2016chaos} that the tripartite information might serve as a quantifier of genuine four-partite entanglement for pure states.

We can get some intuition on $I_3$ by considering its classical counterpart.
We will denote the tripartite information computed with respect to a classical probability distribution $P$ as $J_3(P)\equiv J_3(A:B:C)_P$.
% \cLI{io questa la chiamerei sempre $I_3$ non $J_3$; e' anche la notazione standard penso}
For a balanced probability distribution over the three-bit events $000$ and $111$ we have:
\begin{equation}\label{eq:ex_I3_1}
    p(000) = p(111) = \frac12.
\end{equation}
This gives $J_3=1$.
If any one of the bits is uncorrelated with the others, for example if the distribution is
\begin{equation}\label{eq:ex_I3_0}
    p(001) = p(111) = \frac12,
\end{equation}
then we get $J_3=0$.
Finally, a balanced distribution of the form
\begin{equation}\label{eq:ex_I3_-1}
    p(000) = p(011) = p(101) = p(110) = \frac14,
\end{equation}
gives $J_3=-1$.
We can interpret these results saying that $J_3$ quantifies the redundancy with which one of the bits is encoded in the other two.
In the case of~\cref{eq:ex_I3_1}, the first bit is encoded redundantly in the last two. For~\cref{eq:ex_I3_0}, only one output bit is useful, hence the vanishing redundancy. And for~\cref{eq:ex_I3_-1}, ``negative redundancy'' means that information can only be retrieved by measuring both of the other two bits at the same time, hence there is more information into the pair of bits than there is when they are taken individually.
However, when the mutual information is replaced by the QMI and is applied to generic quantum states, as we will see, this straightforward classical interpretation can get significantly more involved, due to the non-accessible components of the associated QMIs, and the fact that the definition of $I_3$ does not take into account the choice of encoding basis.

\subsection{Non-accessibility of $I_3$}
\label{subsec:accessibility_issues_I3}

% \parTitle{Non-accessibility of QMIs}
One important issue about $I_3$ is that it does not directly quantify correlations between input and output measurement probabilities, owing to its definition via QMIs given in~\cref{eq:definition_tripartite_info}.
In other words, the value of $I_3$ might not be a faithful representation of the amount of information about the input that is lost in correlations between different outputs.
This is due to the well-known fact that the correlations measured by the QMI
% $I(A:B)_\rho\equiv S(\rho_A)+S(\rho_B)-S(\rho)$
are not generally directly accessible via local measurements~\cite{ollivier2001quantum,bera2017quantum,piani2008no,wilde2011classical}. 
Different states can result in identical correlations between measurement outcomes, but nonetheless have different QMIs.
For example, $\rho_1=\PP\left[\frac1{\sqrt2}(|00\rangle+|11\rangle)\right]$ and $\rho_2=\frac12(\mathbb{P}_{00}+\mathbb{P}_{11})$,
both produce fully correlated outcomes when both parties measure in the computational basis, and yet $I(A:B)_{\rho_1}=2 I(A:B)_{\rho_2}=2$.
Here and in the following, we use the shorthand notation
\begin{equation}
    \mathbb{P}(\ket\psi)\equiv 
    % \mathbb{P}_\psi \equiv 
    \ketbra{\psi}{\psi}
\end{equation}
to denote the projection onto a ket state $\ket\psi$.
\addLI{When clear from the context, we will also employ the equivalent notation $\mathbb P_\psi\equiv \mathbb P(|\psi\rangle)$.}

% \parTitle{Non-accessibility of $I_3$}
The QMI not directly quantifying \textit{accessible} correlations~\cite{piani2008no,wilde2011classical} directly propagates into similar issues for $I_3$. For example, one might observe that a fully correlated three-qubit separable state of the form $\rho=\frac12(\mathbb{P}_{000}+\mathbb{P}_{111})$ gives $I_3(\rho)=1$, whereas a GHZ state $\ket{\rm GHZ}\equiv\frac1{\sqrt2}(\ket{000}+\ket{111})$ gives $I_3(\PP_{\rm GHZ})=0$, despite both states translating into an identical degree of retrievability of the information about $A$ from $B$ and $C$. 

% \parTitle{$I_3$ vanishes for pure states}
Another direct way to see that $I_3$ might not be the best quantity to consider is to observe that $I_3(\mathbb{P}_\psi)=0$ for all $\ket\psi$, that is, all pure states correspond to the same value of the tripartite information, even though pure states might correspond to entirely different ways to distribute information among the three parties.
This suggests that the effects of the quantum discord components in $I_3$ are sufficient to conceal important properties of the correlations between different subsystems.

\subsection{Double-arm approach to QIS}
\label{subsec:double-arm}

\begin{figure}[tb]
    \centering
    \includegraphics[scale=0.8]{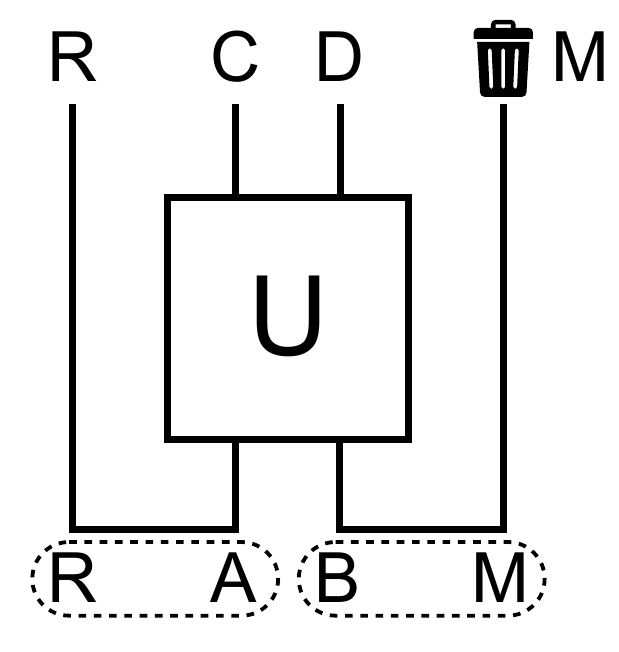}
    \caption{The setup proposed in~\cite{hosur2016chaos} to quantify the QIS of $U$. Maximally entangled states are used to write an output state which encodes correlations between inputs and outputs.
    }
    \label{fig:TMIhosur}
\end{figure}

% \textit{Description of the approach ---}
One of the most popular approaches~\cite{hosur2016chaos} to quantify the QIS of a unitary dynamics%
~\footnote{While we only consider the case of unitary dynamics here, this formalism can be straightforwardly generalized to isometries, by allowing for $\calH_C$ and $\calH_D$ to have different dimensions than $\calH_A$ and $\calH_B$.}
$U:\calH_A\otimes\calH_B\to\calH_C\otimes\calH_D$, where $\calH_A\otimes\calH_B=\calH_C\otimes\calH_D$, involves evaluating the $I_3$ on a tripartite state $\rho_U$ defined as
% \begin{equation}\label{eq:hosur_state}
%     \rho_U\equiv \Tr_D \mathbb{P} \left[
%     (U_{AB}\otimes I_{RM})
%     (\ket*{\Psi^+}_{AR}\otimes\ket*{\Psi^+}_{BM})
%     \right],
% \end{equation}
\begin{equation}\label{eq:hosur_state}
\scalebox{0.9}{$\displaystyle
    \rho_U\equiv \Tr_M \mathbb{P} \left[
    (U_{AB\to CD}\otimes I_{RM})
    (\ket*{\Psi^+}_{AR}\!\otimes\!\ket*{\Psi^+}_{BM})
    \right],
$}
\end{equation}
where $\ket*{\Psi^+}_{XY}$ is a maximally entangled state in the partitions $\calH_X\otimes \calH_Y$.
The state in~\cref{eq:hosur_state} has a simple diagrammatic representation, given in~\cref{fig:TMIhosur}.
The intuition behind this method is to map the channel into a state and to study the QIS properties of the channel through the same entropy-based toolkit we use for states.
It was then argued~\cite{hosur2016chaos} that $I_3(\rho_U)<0$ witnesses the presence of QIS.

% \textit{Why the partial trace? ---}
One way to understand the partial trace used to define $\rho_U$ is to observe that it amounts to probing the properties of the state obtained on the output modes $R,C,D$ (\textit{c.f.}~\cref{fig:TMIhosur}) when we do not know anything about the input used in $B$.
Indeed, the same state $\rho_U$ is obtained by letting the input state $B$ be the maximally mixed state. Moreover, by measuring $M$, the corresponding state conditioned to any measurement outcome is pure, and thus the corresponding $I_3$ zero. This implies that $I_3(\rho_U)$ can be nonzero only if we introduce some effective partial ignorance about the input state.

% \textit{Interpretation via entanglement-assisted capacities ---}
Another issue with the use of maximally entangled states in the definition of $\rho_U$ is that the QMI of a bipartite state of the form $(\Phi\otimes I)\mathbb{P}[\ket*{\Psi^+}]$ quantifies the entanglement-assisted classical transmission rate through the channel $\Phi$~\cite{watrous2018theory}.
That is to say, $I(A:B)_{\rho_\Phi}$ with $\rho_\Phi\equiv(\Phi\otimes I)\mathbb{P}[\ket*{\Psi^+}]$ quantifies how much classical information can be sent through the channel, under the assumption that sender and receiver share the entangled state $\ket*{\Psi^+}$%
~\footnote{More precisely, the entangled state used in the definition of $\rho_U$ is not $\ket*{\Psi^+}$ but rather a tensor product of maximally entangled states on $AR$ and $BM$. This does not however significantly modify our argument here, because the amount of entanglement in the bipartition $RM\otimes CD$ is identical in both cases.}.
It follows that the QMIs in $I_3(\rho_U)$ quantify correlations between input and outputs that are only accessibles assuming these share entangled states, and thus so does $I_3(\rho_U)$.
However, in most contexts where QIS is studied, this is not the case. Rather, one is interested in how much information is recoverable from the outputs, without assuming additional knowledge of correlation between them.

% \parTitle{Tripartite is always nonpositive}
Finally, for any unitary $U$, we have $I_3(\rho_U)\le0$, which implies that the sign of $I_3(\rho_U)$ does not give useful information in such cases.
To see this, given any state of the form~\cref{eq:hosur_state}, observe that the reduced states on $RM$  and $CD$ are always product states, and the entropies satisfy
\begin{equation}\small
\begin{gathered}
    % S(R)=S(\rho_R),\quad
    S(CDR)=S(M), \\
    S(CD) = S(RM) = S(R)+S(M).
    \end{gathered}
\end{equation}
It follows that $I_3$ reads
\begin{equation}
    I_3(CDR) = S(C) + S(D) - S(RC) - S(RD),
\end{equation}
and thus, using the strong subadditivity of the von Neumann entropy, we always have $I_3(CDR)\le0$. A similar argument was presented in~\cite{ding2016conditional}.
This is similar to known results in the context of holography: assuming the Ryu-Takajanagi formula holds~\cite{Ryu:2006bv}, the tripartite information is always negative in gravitational theories with holographic duals~\cite{hayden2013holographic}.
This feature is dubbed ``monogamy of holographic mutual information'' and is considered evidence that quantum gravity should possess some scrambling features.

\parTitle{Example with perfect tensors}
A particularly clear way to see that highly negative values of $I_3$ might not correspond to scrambling dynamics, is to consider a unitary $U:\mathbb{C}^d\otimes\mathbb{C}^d\to \mathbb{C}^d\otimes\mathbb{C}^d$, for some odd $d$, derived from a four-party perfect tensor, of the form~\cite{ding2016conditional}:
\begin{equation}
    U \ket{i}\ket{j} =
    \ket{i+j} \ket{i-j},
\end{equation}
where all arithmetics is modulo $d$.
The corresponding state $\rho_U$ gives $I_3(\rho_U)=-2\log d$, which might lead to the conclusion that $U$ maximally scrambles input information. However, if classical information is encoded in the choice of $\ket i$, fixing the value of $\ket j$, then local measurements can reliably recover the input information.
For example, if we fix $j=0$, the evolution reads
\begin{equation}
    U \ket{i,0} = \ket{i}\ket{i}.
\end{equation}
This shows that there is a way to encode classical information in the input states, such that said information is perfectly recoverable from local measurements of the output.

\subsection{Single-arm approach to QIS}
\label{subsec:single-arm}

\begin{figure}[tb]
    \centering
    \includegraphics[scale=0.8]{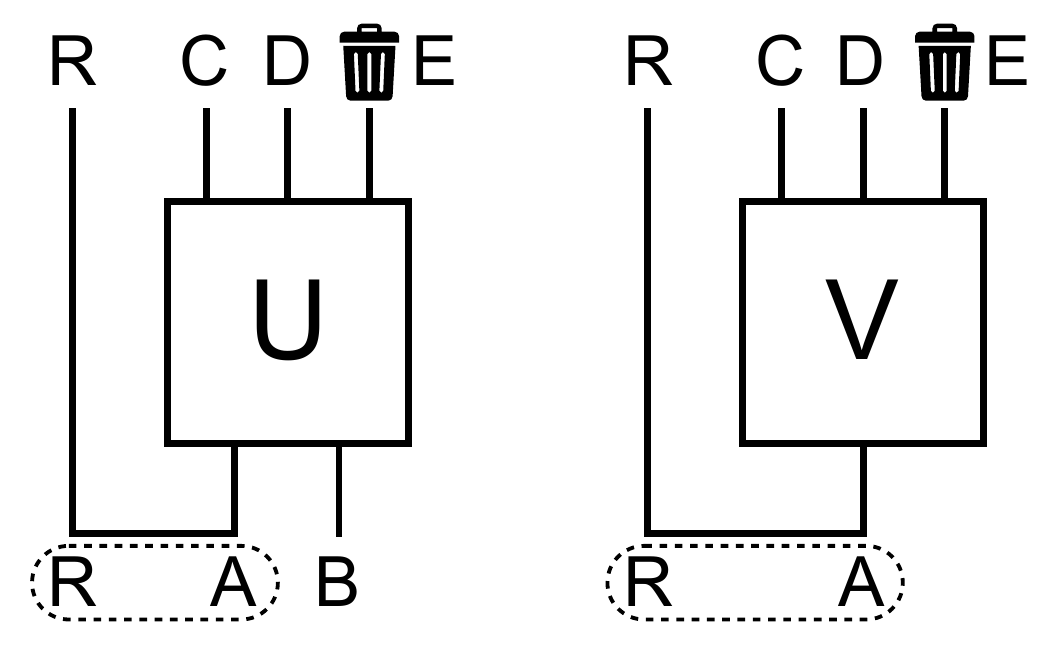}
    \caption{Schematic representations of a possible setup used to test scrambling by means of $I_3$. A register $R$ is initially maximally entangled with the input qubits and keeps track of the stored information. 
    \addLI{In the scheme on the left, input states are injected into the subsystem $A$, while the state in $B$ is assumed to be fixed. Evolving the state in $AB$ through the unitary $U$ is equivalent to the evolution of the state in $A$ through some suitable isometry $V$.}
    }
    \label{fig:TMIpure}
\end{figure}

% \parTitle{Focusing on individual inputs}
An alternative way to use the tripartite information to quantify QIS of a unitary $U$, used in~\cite{swingle2018unscrambling,xu2022scrambling}, is to focus on correlations with respect to a single input system, using a setup like the one in~\cref{fig:TMIpure}.
In this approach, given a unitary $U_{AB\to CDE}:\calH_A\otimes\calH_B\to\calH_C\otimes\calH_D\otimes\calH_E$,
% $U:\calH_A\otimes\calH_B\to\calH_C\otimes\calD$,
the goal is to investigate how information in the input state propagates and is recoverable from different outputs defining a tripartition of the output modes, as shown in~\cref{fig:TMIpure}.
Given the four-partite state
\begin{equation}
\label{eq:SingleArmState}
    \ket{\psi}_{RCDE} \equiv
    (U_{AB\to CDE}\otimes I_R)
    (\ket*{\Psi^+}_{AR}\otimes \ket0_B),
\end{equation}
with $\ket0_B$ some fixed input state in the mode $B$, the QIS is quantified via
\begin{equation}
\label{eq:I3singlearm}
    I_3(R:C:D) \equiv 
    I_3(\Tr_E(\mathbb{P}_{\psi})).
    % I_3(\rho_V),
\end{equation}
The state $\ket\psi_{RCDE}$ can also be equivalently defined via an \textit{isometry} $V_{A\to CDE}$ as
\begin{equation}
    \ket{\psi}_{RCDE} \equiv
    (V_{A\to CDE}\otimes I_R)
    \ket*{\Psi^+}_{AR}.
\end{equation}
\addLI{The corresponding QIS is then defined analogously to the previous case, as
$I_3(R:C:D)=I_3(\rho_V)$ where $\rho_V=\trace_E(\mathbb P_\psi)$.}
%The advantage of this formalism is that we can focus on the properties of the isometry $V\ket\psi\equiv U(\ket\psi\otimes \ket0_B)$, which are the ones of more direct relevance here, rather than study the individual properties of $U$ and $\ket0_B$.
%\cLI{rileggendo qui sembra quasi che intendiamo che ``this approach'' si riferisce al nostro usare isometrie; forse questa frase (prima del commento) possiamo toglierla e basta, visto che diciamo anche prima ``equivalently'' defined?}

However, as we will show through explicit examples, this ``single-arm approach'' to quantifying QIS can also result in misleading values of $I_3$.

\parTitle{Example with $I_3=0$}
Consider the isometry $V:\mathbb{C}^2\to (\mathbb{C}^2)^{\otimes 3}$ defined by
\begin{equation}\label{eq:example_I3_van}\small
\begin{gathered}
    V \ket 0 = \frac{\ket{00}+\ket{11}}{\sqrt2}\ket0,
    \qquad
    V \ket 1 =
    \frac{\ket{01}+\ket{10}}{\sqrt2}\ket0.
\end{gathered}
\end{equation}
The corresponding quantum tripartite information is $I_3(\rho_V)=0$.
On the other hand, the classical probability distribution resulting from encoding information in the computational basis and measuring the first two output qubits, results in a joint distribution identical to~\cref{eq:ex_I3_-1}, which gives $J_3=-1$. Measuring the same two output qubits but encoding information in the basis $\{\ket+,\ket-\}$, the evolution reads
\begin{equation}
    \ket\pm \to \ket{\pm\pm}\ket0,
\end{equation}
meaning that information is locally retrievable from both of the first two output qubits, which is akin to the classical example in~\cref{eq:ex_I3_1}, where instead $J_3=1$.
Furthermore, measuring different combinations of output qubits, we get yet different values of $J_3$: measuring first and third --- or second and third --- output qubits, we get $J_3=0$ for any choice of input encoding.

\parTitle{Example with $I_3=1$}
As another toy example, consider the isometry $V$ acting on single-qubit inputs as
\begin{equation}\label{eq:example_I3_pos}
    V\ket0= \frac{\ket{000}+\ket{111}}{\sqrt2},
    \quad
    V\ket1= \frac{\ket{000}-\ket{111}}{\sqrt2}.
\end{equation}
The associated $\rho_V$ gives $I_3(\rho_V)=1$.
This would seem to indicate that $V$ has no scrambling effects, \textit{i.e.}~input information ends up encoded locally into the output qubits. However, $V$ behaves in exactly the opposite way: unless all three output qubits are measured simultaneously, there is no way to discriminate between the input states $\ket0$ and $\ket1$, meaning that $V$ fully hides such information into the output correlations.
At the same time, we observe that
\begin{equation}
    V\ket+= \ket{000},
    \qquad
    V\ket-= \ket{111},
\end{equation}
which means that the information encoded in the $\{\ket+,\ket-\}$ input basis is encoded in a maximally local way into the output degrees of freedom.
This example shows very clearly that, even though there is a choice of input encoding which gives $J_3=1$, consistently with the result $I_3(\rho_V)=1$, one cannot simply state that $V$ ``does not scramble input information'': whether this happens depends entirely on the way said information has been encoded in the input states.

\parTitle{Example with $I_3=-1$}
Finally, consider the isometry defined as
\begin{equation}\label{eq:example_I3_neg}
\scalebox{1}{$\displaystyle\begin{gathered}
    V\ket0 = \frac{\ket{000}+\ket{111}}{\sqrt2},
    \quad
    V\ket1 = \frac{\ket{001}+\ket{110}}{\sqrt2}.
\end{gathered}$}
\end{equation}
This gives $I_3(\rho_V)=-1$, which should correspond to a high degree of scrambling. This is true when information is encoded in the computational basis. However, observe that in a different basis this isometry gives
\begin{equation}
    V\ket\pm = \left(
    \frac{\ket{00} \pm \ket{11}}{\sqrt2}
    \right)\ket\pm,
\end{equation}
which means that measuring either only the third qubit, or the first two qubits together, allows to recover deterministically the information encoded in the $\{\ket+,\ket-\}$ basis.
This being in contrast with the maximally negative tripartite information,, which would instead suggest $V$ to be maximally scrambling. 

\parTitle{Conclusions}
The examples in~\cref{eq:example_I3_van,eq:example_I3_neg,eq:example_I3_pos} show that the values of $I_3$ need to be interpreted with some degree of caution, as they do not always faithfully represent how much the corresponding dynamics makes information non-retrievable from local measurements of the output.
In the later sections we will show how these issues can be overcome by suitably modifying the definition of QIS in terms of accessible tripartite informations.

\subsection{Encoding-independence of $I_3$}
\addLI{
The examples in \cref{eq:example_I3_van,eq:example_I3_pos,eq:example_I3_neg} show that the amount of information that can be locally retrieved from a subsystem of the output strongly depends on the choice of encoding states. However, the tripartite information evaluated as in~ \cref{eq:hosur_state} or \cref{eq:SingleArmState} is not sensible to the choice of input encoding. In fact, in the setup of~\cref{fig:TMIhosur,fig:TMIpure}, the target unitary $U$ (or isometry $V$) acts on the maximally entangled state $\ket{\Psi}_{RA}$.
This corresponds to assuming complete ignorance about the input state; as a consequence, the triparite information evaluated with this method does not depend on the basis used to encode information in the input states.

\addLI{An explicit way to show this is to observe that because $A$ contains the states fed into the evolution,
applying a local unitary operation on $A$, by replacing $\ket*{\Psi_{AR}^+}$ with $(I_R\otimes U)\ket*{\Psi_{AR}^+}$, corresponds to changing the basis in which information is encoded.
However, a general property of maximally entangled states is that for any local unitary $U$,
}
\begin{equation}
    (I_R\otimes U)\ket*{\Psi_{AR}^+}
    =
    (U^T \otimes I_A)\ket*{\Psi_{AR}^+}.
\end{equation}
\addLI{In other words, changing the input entangled state by a local unitary operation on $A$ is equivalent to changing it applying a local unitary operation on $R$.}
% An explicit way to observe this fact is the following: one could think that a possibility to keep track of the choice of encoding basis is to input a different state of the form:
% \begin{equation}
%     \ket{\tilde\Psi_{RA}}=\mathbb{I}_R\otimes U\ket{\Psi_{RA}}=\frac{1}{\sqrt{|\mathcal{H}_A}|}\sum_{i}\ket{i}_R\otimes U\ket{i}_A,
% \end{equation}
% where $\{\ket{i}_{R,A}\}$ is a basis for $R,A$ and the unitaty $U$ is a change of basis in the input space $A$. However, the maximally entangled state has the property that:
% \begin{equation}
%     \ket{\tilde\Psi_{RA}}=U^{T}\otimes{\mathbb{I}}_A \ket{\Psi_{RA}}\,.
% \end{equation}
Since $R$ is an auxiliary space and the choice of basis for it does not affect the value of $I_3$, this implies that the tripartite information is encoding-independent, as previously claimed.
}
\addLI{Note that, as we will show in the later sections, defining QIS via accessible mutual informations, does instead allow to naturally capture the importance of the input encoding basis.}

\section{Accessible tripartite information}
\label{sec:accessible_I3}

% \parTitle{Summary of section}
In this section we propose a way to overcome the critical aspects of the use of $I_3$ as a scrambling quantifier, discussed in~\cref{sec:issues_I3}.
We show that, rather than QMIs, the quantities of most direct relevance are the classical capacities of channels induced by the unitaries under examination, and more specifically the accessible informations associated to such channels.
Furthermore, the accessible information reveals how different output subsystems might be maximally correlated with respect to different encoding bases. This is a fundamental feature of QIS that, to our knowledge, has not been explored so far.

\parTitle{Outline}
We start discussing in~\cref{subsec:MI_vs_capacities} the close relation between quantifying scrambling and quantifying (specific types of) classical capacities of channels derived from the target evolution.
We then introduce in~\cref{subsec:accessible_tripartite} our definition of ``accessible tripartite information'', and discuss how it compares with the standard tripartite information in a number of examples.

\subsection{QIS in terms of channel capacities}
\label{subsec:MI_vs_capacities}

% \parTitle{Why consider capacities instead}
Many of the issues discussed in~\cref{sec:issues_I3} are associated with the non-accessibility of QMIs.
These can be avoided by using instead accessible informations~\cite{shor2003capacities,wilde2011classical}.
More precisely, since the goal is quantifying the correlations between an input $\ket\psi$ and the information retrievable from a subset $S$ of the output system $V\ket\psi$, it follows that the most relevant quantities are the capacities~\cite{watrous2018theory,wilde2011classical,shor2003capacities} of the channel $\Phi_S$ defined as
\begin{equation}
\label{eq:channel_trace}
    \Phi_S(\rho)\equiv\Tr_{\bar S}(V \rho V^\dagger),
\end{equation}
where $\Tr_{\bar S}$ denotes the partial trace with respect to all degrees of freedom except for the ones in $S$.
Here $V$ is a generic isometry like the ones introduced in \autoref{subsec:single-arm}.
Different kinds of channel capacities quantify achievable rates of classical and quantum information transmission.
In the context of QIS, it is not generally clear which type of information is the most relevant.
Here we focus on quantifying the amount of information transmissible via a given dynamics and recoverable via local measurements.

% \parTitle{Classical vs accessible informations}
However, the classical capacity of a channel quantifies the rate of information transmission achievable over many uses of the channel, when sender and receiver are allowed to use entangled input states and global measurements on the corresponding outcomes.
In the context of QIS, this might not be the most appropriate quantity to focus on: both if we think of QIS as the amount of information retrievable about input states, and if we are interested in QIS from the perspective of the black-hole information paradox, considering correlations that are only accessible using input states that are entangled across multiple injections in the channel may not be the relevant figure of merit.
Rather, we have to consider correlations accessible without the use of entanglement between sender and receiver, and without nonlocal measurements.
The associated quantity is sometimes referred to as the ``accessible information of the channel'', or $C_{1,1}$ capacity~\cite{shor2003capacities,wilde2011classical}.

\parTitle{Accessible information of a channel}
Given a bipartite state $\rho$, its \textit{accessible information} is the classical mutual information of the joint probability distribution corresponding to a choice of local measurements, maximized over all possible such choices.
That is,
\begin{equation}\label{eq:definition_accessible_info}
    I_{\rm acc}(\rho) \equiv \sup_{\mu^A,\mu^B} I(A:B)_p,
\end{equation}
where the classical probability distribution $p$ is given by
\begin{equation}
    p(a,b) = \Tr[(\mu^A_a\otimes\mu^B_b)\rho],
\end{equation}
and we maximize over the possible local POVMs $\{\mu_a^A\}$ and $\{\mu^B_b\}$.
\addLI{On the other hand, the accessible information of a \textit{channel} $\Phi$ is the maximal accessible information obtainable varying over all possible ways to encode information at the input, and all possible ways to measure the output state to retrieve.
This can be written formally as the accessible information of states of the form $(I_R\otimes\Phi)\rho_{cq}$, maximized over all possible classical-quantum states $\rho_{cq}$.
Note that classical-quantum states are those separable states admitting a decomposition of the form
\begin{equation}
    \rho_{cq}=\sum_x p_x \mathbb{P}_x\otimes \rho_x
\end{equation}
for some input ensemble $(p_x,\rho_x)$, and with $\mathbb{P}_x$ projections over an orthonormal basis for the auxiliary space $R$.
The introduction of this auxiliary space $R$ is only a formal tool to build into the state the knowledge that the different states $\rho_x$ refer to states fed into the channel at different times.}
% On the other hand, the accessible information of a \textit{channel} $\Phi$ is the maximal accessible information of states of the form $(I\otimes\Phi)\rho_{cq}$ for all classical-quantum states $\rho_{cq}$:
\addLI{In summary, we define the accessible information of a channel $\Phi$ as
\begin{equation}\label{eq:def_accessible_info_channel}
     I_{\rm acc}(\Phi) \equiv
     \sup_{\rho_{cq}} I_{\rm acc}((I_R\otimes\Phi)\rho_{cq}).
\end{equation}}
% \addG{where the classical-quantum state can be written as $\rho_{cq} = \sum_x p_x \PP_x\otimes \rho_x$. Here $\mathbb{P}_x$ represents a computational basis for an auxiliary register $X$ and $\rho_x$ is the input state living in the space $A$.
% }.
\addLI{We can note the tight connection between this accessible information in~\eqref{eq:def_accessible_info_channel} and the QMI, we can also write it as
\begin{equation}
    I_{\rm acc}(\Phi) =
    \sup_{\rho_{cq},\mu} 
    I\left(
    (I_R\otimes \Delta_\mu)
    (I_R\otimes \Phi) \rho_{cq}
    \right),
\end{equation}
where $\Delta_\mu(\rho)\equiv \sum_y \trace(\mu_y\rho)\,\mathbb{P}_y$ is the channel describing the decoherence induced by measuring with the POVM $\mu$.
It is also worth observing that if we remove the measurement decoherence $\Delta_\mu$, and allow for arbitrary input states $\rho$, we recover a definition analogous to \eqref{eq:I3singlearm}.
}
%\cLI{this is probably enough}
%of the classical-classical state
%\begin{equation}
%    (I\otimes \Delta_\mu)\left(
%    \sum_x p_x \mathbb{P}_x\otimes \Phi(\rho_x)
 %   \right),
%\end{equation}
%maximized over all ensembles $\{(p_x,\rho_x)\}_x$ and POVMs $\mu$, where we used $\Delta_\mu$ to denote the entanglement-breaking channel~\cite{watrous2018theory} $\Delta_\mu(\rho)\equiv\sum_y \Tr(\mu_y \rho) \mathbb{P}_y$.
%\cLI{forse ora questa roba si puo togliere, ho spiegato prima della 26}

\addLI{In the following, the accessible information will be always understood as evaluated on the evolved classical-quantum state $\tilde{\rho}=I_R\otimes\Phi_{A\rightarrow S\overline{S}}(\rho_{cq})$ :
\begin{equation}
\label{eq:notation_Iacc}
I_{\text{acc}}(R:S)_{\tilde{\rho}}=I_{\text{acc}}(\text{Tr}_{\overline{S}}\,\tilde{\rho})\,.
\end{equation}
%\cLI{questa notazione confonde anche me =)}
Whenever the subscript $\Phi(\rho_{cq})$ is omitted, $I_\text{acc}(R:S)$ is understood as optimized over the choice of classical-quantum state $\rho_{cq}$. %Let us also stress that it would be more correct to denote the accessible information in \eqref{eq:notation_Iacc} as $I_{\text{acc}}(X:S)_{\Phi(\rho_{cq})}$. With a small abuse of notation, in the following we will denote with the same symbol $A$ both the classical register and the input space, if not ambigous.
}

%\addLI{\redsout{When more registers are involved but we need to compute the mutual information between only a pair of them, for example if $\rho$ lives in a space $\calH_A\otimes\calH_B\otimes\calH_C$ but we need the mutual information between $A$ and $B$, we use the equivalent notation}
%\begin{equation}
%    I_{\rm acc}(A:B)_\rho \equiv
%    I_{\rm acc}(\rho_{AB}),
%\end{equation}
%\redsout{and similarly for any other pair of relevant registers.}
%}

This latter formulation, in particular, shows the similarities, and differences, with the way the QMI was calculated in~\cref{subsec:single-arm,subsec:double-arm}.
We can also consider the mutual information in~\cref{eq:def_accessible_info_channel} for fixed choices of input ensembles, which tells us the maximum rate of information transmission that can be substained by the channel for a fixed input encoding, whose letters are the states $\rho_x$ occurring with probabilities $p_x$.
Computing classical capacities --- and in particular $C_{1,1}$ capacities --- is in general a daunting task, due to the maximization required for the calculation.
We focus here on examples where this calculation can be performed, to illustrate the differences that can incur from considering these types of quantities, at least in principle.

\subsection{Accessible tripartite informations}
\label{subsec:accessible_tripartite}

% \parTitle{Definition of accessible $I_3$}
Following our discussion on the use of accessible mutual informations, a natural proposal for a measurement of quantum information scrambling more tightly related to what we seek to quantify is to define the \textit{accessible tripartite information}
$I_{3}^{\rm acc}(R:C:D)$.
\addLI{
More formally, we define this quantity as}
\begin{equation}\label{eq:acc_tripartite}
    % \small
    % I_{3,\rm acc}(\rho)\equiv
    I^{\rm acc}_3(R\!:\!C\!:\!D)=
    I_{\rm acc}(R\!:\!C) +
    I_{\rm acc}(R\!:\!D) -
    I_{\rm acc}(R\!:\!CD),
\end{equation}
\addLI{for a given tripartition $CDE$ of the output and evolution $\Phi_{A\rightarrow CDE}$. The subspace $E$ could be actually empty, {\it i.e.} the output space is actually bipatitioned; in this case, it turns out $I_{\rm acc}(R\!:\!CD)=\log_2|\calH_R|$ for isometric evolutions, since the input information can be always retrieved performing an appropriate global measurement. 
Each summand in \eqref{eq:acc_tripartite} is independently  optimized over the classical-quantum states $\rho_{cq}$, {\it i.e.} over the choice of input states (c.f.r \Cref{eq:notation_Iacc}).
%\addLI{where $\rho_x$ represents the states living in $A$ and used at the input, and $\Phi(\rho_x)$ live in the larger space $CDE$. Here $\PP_x$ is a computational basis for the auxiliary classical register $R$, used to compute the accessible information.
%This choice allows to fix a common encoding basis for the three terms in~\cref{eq:acc_tripartite}.
% It is worth noting that $\rho$ does not represent a physical state.
% It is worth noting that the state $\rho$ is used to describe the physical process of evolving the states $\rho_x$ through $\Phi$.
In this definition, each of the three accessible information will also require a separate optimization over the output bases.
}

% Each term in this expression will require to perform an optimization over both input and output measurement choices, and $\rho$ is here the classical-quantum state
% \begin{equation}
%     \rho = \sum_x p_x \PP_x\otimes \Phi(\rho_x),
% \end{equation}
% where the choice of input ensemble $\{(p_x,\rho_x)\}$ also needs to be maximized.
% \addLI{The computation of~\cref{eq:acc_tripartite} involves choosing an input encoding basis, via the choice of $\rho$, and optimizing over the possible measurement choices at the output. In this definition, each of the three accessible information will require a separate optimization over the output bases.}

\begin{figure}[tb]
    \centering
    \includegraphics[scale=0.8]{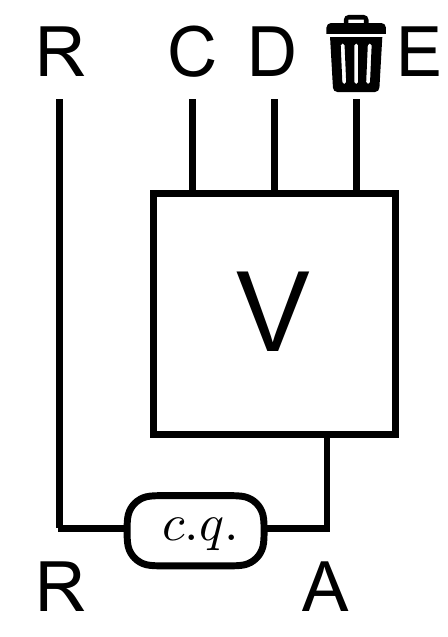}
    \caption{ Schematical representation of the setup proposed in section \ref{subsec:accessible_tripartite} in order to estimate information scrambling by means of the accessible tripartite information. A classical-quantum state $\rho_{AR}$ is prepared, $R$ being a register keeping track of the input information. The input qubits live in $A$ and evolve under the action of the target isometry $V$.
    }
    \label{fig:TMIacc}
\end{figure}

Our proposal is to compute $I_{3}^{\rm acc}(R:C:D)$, where $R$ represents the classical register, and $C$ and $D$ a pair of output subsystems --- which might in general encompass more than one qubit each --- in order to understand how much information about $R$ is hidden in correlations between $C$ and $D$. The choice of output subspaces $C$ and $D$ will dramatically affect the corresponding value of $I_{3}^{\rm acc}$, which reflects the fact that information might be encoded in correlations among a specific subset of the output qubits. Figure \ref{fig:TMIacc} provides a sketch of the proposed setup.

% \parTitle{Properties of accessible $I_3$}
The accessible tripartite information quantifies how much more correlations are present between the input register $R$ and the output $C$ and $D$, rather than between $R$ and $CD$ taken together.
This quantity can now therefore be interpreted as a direct analogue of the classical tripartite information discussed in~\cref{subsec:definition_of_I3}.
In particular, this also means that $I_{3}^{\rm acc}$ quantifies how much information about $A$ is \textit{hidden} in the correlations between $C$ and $D$, or in other words, how much more information about $A$ is accessible when one can measure $C$ and $D$ together rather than separately. In some circumstances, this does not tell us how much information about inputs is recoverable from outputs: for example, $I_{3}^{\rm acc}(A:C:D)=0$ could both mean that there is no correlation at all between inputs and outputs, or that there is maximal correlation, but the input information can be retrieved completely via local measurements.
Depending on the circumstances, one might thus be more interested in studying accessible informations between input and individual outputs, rather than tripartite informations. We leave a more detalied analysis of this perspective for future work.

\subsection{Examples}

\parTitle{Example with $I_3=0$}
Consider as an explicit example the isometry given in~\cref{eq:example_I3_van}.
% , whose action reads:
% \begin{equation}
%     V\ket0=\frac{\ket{00}+\ket{11}}{\sqrt2} \ket0,
%     \quad
%     V\ket1=\frac{\ket{01}+\ket{10}}{\sqrt2} \ket0.
% \end{equation}
While the quantum tripartite information would give $I_3=0$ in this case, using accessible tripartite informations we get different values depending on the partition chosen for the output space.
We have
\begin{equation}
\begin{gathered}
    I_{\rm acc}(R:1) = I_{\rm acc}(R:2) =1,\,\quad I_{\rm acc}(R:3) = 0, \\
    I_{\rm acc}(R:12)=
    I_{\rm acc}(R:13) = I_{\rm acc}(R:23) = 1,
\end{gathered}
\end{equation}
and thus
\begin{equation}
\begin{gathered}
    I_3^{\rm acc}(R\!:\!1\!:\!2) {=} 1,\;\;
    I_3^{\rm acc}(R\!:\!1\!:\!3) {=} I_3^{\rm acc}(R\!:\!2\!:\!3) {=} 0 \\
    I_3^{\rm acc}(R\!:\!1\!:\!23) {=} I_3^{\rm acc}(R\!:\!2\!:\!13) {=} 1,\;
    I_3^{\rm acc}(R\!:\!12\!:\!3) {=} 0
\end{gathered}
\end{equation}
% \addRev{\begin{equation}
% \begin{gathered}
%     I_3^{\rm acc}(1:2) = -1,
%     \,\,
%     I_3^{\rm acc}(1:3) = I_3^{\rm acc}(2:3) = 0, \\
%     I_3^{\rm acc}(1:23) = I_3^{\rm acc}(2:13) = -1,
%     \,\,
%     I_3^{\rm acc}(12:3) = 0.
% \end{gathered}
% \end{equation}}
These reflect how input qubits are encoded in the output space in a way that is partially nonlocal.
Note how, using the fully quantum tripartite information $I_3$ discussed in~\cref{subsec:double-arm}, we only obtain a single quantity, which cannot capture the complexity of the information encoding we get even in this extremely simple toy example.

\parTitle{Example with $I_3=1$}
In this case, all accessible informations are equal to $1$, corresponding to the fact that the input basis $\ket\pm$ is transmitted with maximal redundancy on the output states.
The accessible tripartite informations are correspondingly also all maximal, showing that this isometry is as non-scrambling as one can be.
More explicitly:
\cref{eq:example_I3_pos}
\begin{equation}
\begin{gathered}
    I_{\rm acc}(R:1) = I_{\rm acc}(R:2) = I_{\rm acc}(R:3) = 1, \\
    I_{\rm acc}(R:12) =
    I_{\rm acc}(R:13) =
    I_{\rm acc}(R:23) = 1.
\end{gathered}
\end{equation}
In this case all accessible tripartite informations are $I_3^{\rm acc}(R:i:j)=1$.
% and the accessible tripartite informations read
% \begin{equation}
% \begin{gathered}
%     I_3^{\rm acc}(A:2:3)=I_3^{\rm acc}(A:1:3)=0,
%     \\
%     I_3^{\rm acc}(A:1:2)=-1, \\
%     I_3^{\rm acc}(A:1:23) = I_3^{\rm acc}(A:2:13) = 0,
%     \\
%     I_3^{\rm acc}(A:12:3) = 1.
% \end{gathered}
% \end{equation}

\parTitle{Example with $I_3=-1$}
Let us now consider again the toy example in~\cref{eq:example_I3_neg}, which gave $I_3=-1$.
The individual accessible informations are
\begin{equation}
\begin{gathered}
    I_{\rm acc}(R:1) = I_{\rm acc}(R:2) = 0,
    \quad
    I_{\rm acc}(R:3) = 1,
    \\
    I_{\rm acc}(R:12) =
    I_{\rm acc}(R:13) =
    I_{\rm acc}(R:23) = 1,
\end{gathered}
\end{equation}
and the accessible tripartite informations read
\begin{equation}
\begin{gathered}
    I_3^{\rm acc}(R\!:\!1\!:\!2){=}{-}1, \;\;
    I_3^{\rm acc}(R\!:\!2\!:\!3){=}I_3^{\rm acc}(R\!:\!1\!:\!3){=}0\\
    I_3^{\rm acc}(R\!:\!1\!:\!23) {=} I_3^{\rm acc}(R\!:\!2\!:\!13) {=} 0,\,\,
    I_3^{\rm acc}(R\!:\!12\!:\!3) {=} 1
\end{gathered}
\end{equation}
% \begin{equation}
% \begin{gathered}
%     I_3^{\rm acc}(2:3)=I_3^{\rm acc}(1:3)=0,
%     \quad
%     I_3^{\rm acc}(1:2)=-1, \\
%     I_3^{\rm acc}(1:23) = I_3^{\rm acc}(2:13) = 0,
%     \,\,
%     I_3^{\rm acc}(12:3) = 1.
% \end{gathered}
% \end{equation}
These results are consistent with the information about input states being encoded redundantly in both third and first two qubits.
Note that these values of $I_3^{\rm acc}$ are not all independent from one another: for example, $I_3^{\rm acc}(R:12:3)=1$ and $I_3^{\rm acc}(R:1:3)=I_3^{\rm acc}(R:2:3)=0$ tell us that the input is encoded in both third and first two qubits, but that measuring only second or third qubits individually we do not find the same redundancy; it must then follow that the encoding in the first two qubits is nonlocal, as then verified by $I_3^{\rm acc}(R:1:2)=-1$.

\section{Encoding bases}
\label{sec:encoding_bases}

% \parTitle{Consequences of maximal accessible informations}
In all the examples we considered in~\cref{subsec:accessible_tripartite}, we had $I_{\rm acc}(R:S)=1$ for some subset of output qubits $S$.
Let us assume, without loss of generality, that we achieve this value using the computational basis for encoding.
The saturation of the accessible information has an interesting consequence: regardless of the value of the accessible information $I_{\rm acc}(R:\bar S)$, where $\bar S$ corresponds to all the qubits not in $S$, if the input is encoded in a basis unbiased with respect to the computational one, then its correlations with the state in $\bar S$ must be zero.
This is a direct consequence of Holevo's theorem: if there is full correlation between the input in a basis and some output, and any non-vanishing correlation between input in a different basis and some other piece of output, then it would be possible to acquire information about the input state in more than one basis at the same time.

% \parTitle{Accessible informations in different bases}
The fact that in the shown examples $I_{\rm acc}(R:S)=1$ for some $S$, allowed us to avoid additional complications arising from the previously given definition of accessible tripartite information.
Namely, the fact that the accessible informations that appear in the definition in \cref{eq:acc_tripartite} might be realized by different input bases.

\parTitle{$W_3$ isometry and induced channels}
As an example, consider the isometry $V_{W_3}$ defined by
\begin{equation}\label{eq:W3_isometry}
\begin{gathered}
    V_{W_3}\ket0 = \frac1{\sqrt3}(\ket{100}+\ket{010} + \ket{001}), \\
    V_{W_3}\ket1 = \frac1{\sqrt3}(\ket{100}+\omega \ket{010} + \omega^2\ket{001}),
\end{gathered}
\end{equation}
where $\omega\equiv e^{2\pi i/3}$, so that $1+\omega+\omega^2=0$ and $\bar\omega=\omega^2$.
Consider the channel corresponding to this isometry when looking only at the first output qubit after evolution: $\Phi_V^{(1)}(\rho)\equiv \Tr_{2,3}(V\rho V^\dagger)$. Given a generic pure state $\ket\psi\equiv \sqrt{p_0}\ket0+\sqrt{p_1}e^{i\phi}\ket1$, this reads:
\begin{equation}
    \Phi_V^{(1)}(\mathbb{P}_\psi) =
    \frac23 \mathbb{P}_0 + \frac13 \mathbb{P}_1 
    - \frac23 \sqrt{p_0 p_1}\cos(\phi) Z,
\end{equation}
with $Z\equiv \mathbb{P}_0-\mathbb{P}_1$ the Pauli $Z$ matrix. On the other hand, looking at second or third output qubits we get
% \begin{equation}
%     \Phi_V^{(2)}(\mathbb{P}(\alpha\ket0+\beta\ket1)) =
%     \frac23 \mathbb{P}_0 + \frac13 \mathbb{P}_1 
%     + \frac13 (\Re(\bar\alpha\beta)+\sqrt3\Im(\bar\alpha\beta)) Z,
% \end{equation}
\begin{equation}
\begin{gathered}
    \Phi_V^{(2)}(\mathbb{P}_\psi) =
    \frac13\begin{pmatrix}
        2 + \sqrt{p_0 p_1} S_+ & 0\\
        0 & 1 - \sqrt{p_0 p_1} S_+
    \end{pmatrix}, \\
    \Phi_V^{(3)}(\mathbb{P}_\psi) =
    \frac13\begin{pmatrix}
        2 + \sqrt{p_0 p_1} S_- & 0\\
        0 & 1 - \sqrt{p_0 p_1} S_-
    \end{pmatrix}
\end{gathered}
\end{equation}
where $S_\pm \equiv \cos(\phi)\pm \sqrt3 \sin(\phi)$.

\begin{table*}[tbp]
    \centering
    \setlength{\tabcolsep}{7pt}
    \hypersetup{
    colorlinks=true,
    linkcolor=MidnightBlue,
    filecolor=magenta,      
    urlcolor=cyan
    }
    \begin{tabular}
    {cccccccc}
    \toprule
         $V$ &  $I_3(R{:}i{:}j)$ & $I_3^{\rm acc}(R{:}1{:}2)$ & $I_3^{\rm acc}(R{:}1{:}3)$ & $I_3^{\rm acc}(R{:}2{:}3)$ & $J_3^{\rm acc}(R{:}1{:}2)$ & $J_3^{\rm acc}(R{:}1{:}3)$ & $J_3^{\rm acc}(R{:}2{:}3)$
        \\
        \noalign{\smallskip}
        \hline
        \noalign{\smallskip}
        \multirow{1}{*}{\cref{eq:example_I3_van}}  
        & 0 & 1 & 0 & 0 & 1 & 0 & 0\\
        \noalign{\smallskip}
        \hline
        \noalign{\smallskip}
        \multirow{1}{*}{\cref{eq:example_I3_pos}} 
        & 1 & 1 & 1 & 1 & 1 & 1 & 1\\ 
        \noalign{\smallskip}
        \hline
        \noalign{\smallskip}
        \multirow{1}{*}{\cref{eq:example_I3_neg}} 
        & -1 & -1 & 0 & 0 & 0 & 0 & 0 \\ 
        \noalign{\smallskip}
        \hline
        %\noalign{\smallskip}
        %\multirow{1}{*}{\cref{eq:W3_isometry}} 
        %& ${-}\frac{3}{2}\log _2\left(\frac{4}{3}\right)$ & $2\log_2\left(\frac{2{+}3 \sqrt{3}}{3 \sqrt{3}}\right){-}1$ & $2\log _2\left(1{+}\frac{2}{3 \sqrt{3}}\right){-}1$ & $2\log _2\left(1{+}\frac{2}{3 \sqrt{3}}\right){-}1$ &  \\
        %\noalign{\smallskip}
        %\hline
        \noalign{\smallskip}
        \multirow{1}{*}{\cref{eq:W3_isometry}} 
        & {-}0.623 & {-}0.060 & {-}0.060 & {-}0.060 & 0.186 & 0.186 & 0.186 \\
        \noalign{\smallskip}
    \bottomrule
    \end{tabular}
    \caption{
    \addLI{A comparison between the discussed tripartite-information-based quantifiers of QIS, for all the isometries discussed.
    Note that $I_3$ is independent on the tripartition of the output in all the discussed examples. The accessible tripartition informations $I_3^{\rm acc}$ and $J_3^{\rm acc}$ depend instead on the choice of tripartition in the case of the isometries \cref{eq:example_I3_van} and \cref{eq:example_I3_neg}
    }
    % \textcolor{red}{A comparison among the amount of information scrambling estimated through the various definitions of tripartite information discussed in the main text, for all the isometries considered. The tripartite information $I_3$ is the same independently on the tripartition of the output in all the discussed examples. The accessible tripartition information $I_3^{\rm acc}$ and $J_3^{\rm acc}$ depend instead on the choice of tripartition in the case of the isometries \cref{eq:example_I3_van} and \cref{eq:example_I3_neg}.}
    }
\label{table:summary_of_eqs}
\end{table*}

\parTitle{Optimal encoding bases}
The encoding basis corresponding to maximum information transmission on the first output qubit is $\{\ket+,\ket-\}$. This follows from
\begin{equation}
    \Phi_V^{(1)}(\mathbb{P}_\psi) -\Phi_V^{(1)}(\mathbb{P}_{\psi_\perp}) =
    -\frac{4}{3}\sqrt{p_0 p_1}\cos(\phi) Z,
\end{equation}
implying that the trace distance, and thus the probability of discriminating between the two states, is maximized for $\phi=0$ and $p_0=p_1$.
% can be obesrved maximizing the trace distance $\|\Phi_V^{(1)}(\mathbb{P}_\psi) -\Phi_V^{(1)}(\mathbb{P}_{\psi_\perp})\|_1$ among all pairs of orthonormal qubit states $\ket\psi$ and $\ket{\psi_\perp}$.
On the other hand, for second and third output qubits, we have
\begin{equation}
\begin{gathered}
    \Phi_V^{(2)}(\mathbb{P}_\psi) -\Phi_V^{(2)}(\mathbb{P}_{\psi_\perp}) =
    \frac{2}{3}\sqrt{p_0 p_1} S_+ Z, \\
    \Phi_V^{(3)}(\mathbb{P}_\psi) -\Phi_V^{(3)}(\mathbb{P}_{\psi_\perp}) =
    \frac{2}{3}\sqrt{p_0 p_1} S_- Z,
\end{gathered}
\end{equation}
and the corresponding trace distances are therefore maximized for balanced states with $\phi=\pi/3$ and $\phi=2\pi/3$, respectively.
It follows that the optimal encoding basis is $\left\{\frac{1}{\sqrt2}(\ket0-\omega^2 \ket1),\frac{1}{\sqrt2}(\ket0+\omega^2 \ket1)\right\}$ for the second qubit, and $\left\{\frac{1}{\sqrt2}(\ket0+\omega \ket1),\frac{1}{\sqrt2}(\ket0-\omega \ket1)\right\}$ for the third one.

These results do not contradict our previous remarks about encoding bases, because in all three cases the mutual information corresponding to the optimal basis is not maximal.
Finally, observe that looking at the first two output qubits, $\mathbb{P}_\psi$ evolves to
\begin{equation}\small
    \frac13
    |\alpha+\beta\omega^2|^2 \,\mathbb{P}_{00} +
    \frac13\mathbb{P}\big(
        (\alpha+\beta)\ket{10}+(\alpha+\beta\omega)\ket{01}
    \big).
    % \frac13\begin{pmatrix}
        % |\alpha+\beta\omega^2|^2 & 0 & 0 & 0
    % \end{pmatrix}
\end{equation}
It follows that the orthogonal states $\{\ket0+\omega\ket1, \ket0-\omega\ket1\}$ are sent to two orthogonal states in the output, meaning that the accessible information is maximal when the first two output qubits are measured.
%\cLI{someone double check these please}

\parTitle{Tripartite accessible information with independent optimizations}
To obtain the associated accessible informations, we proceed by computing the conditional probability distributions associated with measuring the various output qubits, using optimal encoding and measuring bases in each case.
For the first output qubit, we encode in the basis $\{\ket+,\ket-\}$ and measure in the computational basis.
The resulting conditional distribution reads:
\begin{equation}\scalebox{0.9}{$\displaystyle
    P(0|+) = \frac13,
    \,\, P(1|+) = \frac23,
    \,\, P(0|-)=1,
    \,\, P(1|-)=0.
$}\end{equation}
This can be recognized as a binary asymmetric channel, also referred to as a $Z$ channel~\cite{mackay2003information}, and has capacity $\log_2\left(1+\frac{2}{3\sqrt3}\right)$.
The same capacity is obtained measuring only second or only third qubit.
Note that these capacities are achieved with different weights used for the input encoding.
%\cLI{not sure how clear this is, but I'm also not sure how much more in detail I should go about this process. To explain it properly we should explain how capacities work in general}
On the other hand, measuring any pair of output qubits simultaneously, the corresponding conditional probability distribution has unit capacity. We conclude that computing the equivalent of the accessible tripartite informations with respect to some pair of output qubits, would then be
% \cD{In light of what we say later, I would be more inclined not to call this quantity accessible tripartite, and would instead use a more cautious/ambiguous terminology.}
% \cLI{ti direi pure di si, ma poi dobbiamo cambiare la terminologia anche in tutti i pezzi prima}
\begin{equation}
    I_3^{\rm acc}(R:1:2)
    = 2\log_2\left(1+\frac{2}{3\sqrt3}\right) - 1
    \simeq -0.06. 
\end{equation}

\begin{figure}[tbh]
    \centering
    \includegraphics[width=\linewidth]{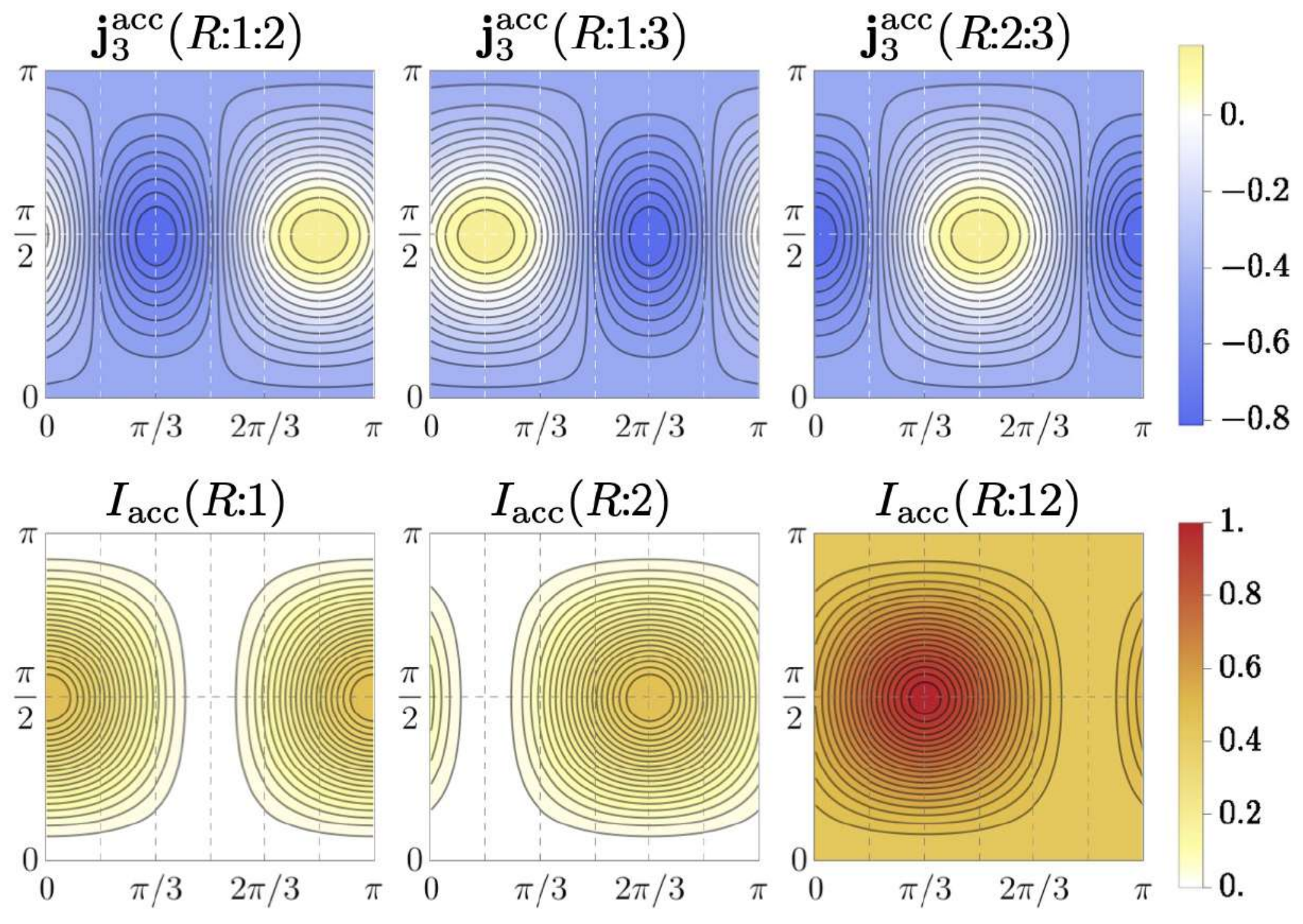}
    \caption{
    Encoding-dependent tripartite and mutual informations corresponding to the isometry $V_{W_3}$ defined in~\cref{eq:W3_isometry}.
    (\textbf{\textit{top row}}) The tripartite informations $\mathbf{j}_3^{\rm acc}(R\!:\!1\!:\!2),\mathbf{j}_3^{\rm acc}(R\!:\!1\!:\!3),\mathbf{j}_3^{\rm acc}(R\!:\!2\!:\!3)$, respectively.
    These are calculated fixing a common input encoding basis for all three constituents accessible informations.
    (\textbf{\textit{bottom row}}) The accessible informations that make up the encoding-dependent information $\mathbf{j}_3^{\rm acc}(R\!:\!1\!:\!2)$. From left to right, these are $I_{\rm acc}(R:1),\,I_{\rm acc}(R:2),\,I_{\rm acc}(R:12)$, respectively.
    }
    \label{fig:accessible_tripartite_infos_W3}
\end{figure}

\parTitle{Better tripartite accessible information}
The fact that the optimal encoding basis changes depending on which output qubits are measured, however, implies that computing $I_3^{\rm acc}$ by simply summing together individual accessible informations is a potentially problematic way to quantify how information in a given encoding basis is spread among different output subsystems: one might find, like we have in this example, that $I_{\rm acc}(R\!:\!C)$, $I_{\rm acc}(R\!:\!D)$ and $I_{\rm acc}(R\!:\!CD)$ all quantify correlations with respect to different encoding bases.
In such cases, a modified tripartite information can be defined, where the maximization with respect to the input basis is performed simultaneously on all its component factors. %{\color{red} I would remove all this part:
% \begin{equation}\scalebox{0.84}{$\displaystyle
%     J_3^{\rm acc}(1:2) \equiv
%     \sup_\Pi [
%     I_{\rm acc}(i_\Pi:1) +
%     I_{\rm acc}(i_\Pi:2) -
%     I_{\rm acc}(i_\Pi:12)
%     ],
% $}\end{equation}
%\begin{equation}
%\label{eq:J3_first}
%\begin{aligned}
%\small
%    J_3^{\rm acc}(A\!:\!1\!:\!2) \equiv
%    \sup_\Pi [
%    I_{\rm acc}(A_\Pi:C) &+
%    I_{\rm acc}(A_\Pi:D) \\
%    &- I_{\rm acc}(A_\Pi:CD)
%    ],
%\end{aligned}
%\end{equation}
%where $I_{\rm acc}(A_\Pi:B)$ denotes the accessible mutual information with the input states measured in the basis $\Pi\equiv \{\Pi_0,\Pi_1\}$.
%More precisely, given an arbitrary classical-quantum state $\rho_{AB}$ of the form
%\begin{equation}
%    \rho_{AB} = p_0 \mathbb{P}_0\otimes\rho_0+p_1\mathbb{P}_1\otimes \rho_1,
%\end{equation}
%we need to compute the associated evolved state
%\begin{equation}   \tilde{\rho}_{ABC}\equiv (I\otimes\Phi)(\rho_{AB})
%    = \sum_i p_i \mathbb{P}_i \otimes \Phi(\rho_i).
%\end{equation}
%\addRev{When clear from the context, we will also use the shorthand notation $\tilde{\rho}\equiv\tilde{\rho}_{ABC}$.}
%where $\Phi\equiv\Phi_{B\to BC}$ is the (in our cases isometric) evolution under consideration.}
More precisely, given an input classical-quantum state of the form $\rho_{cq}=\sum_x p_x\,\PP_x\otimes \rho_x$,
we compute the encoding-dependent tripartite information of the evolved state $\tilde{\rho}=\Phi(\rho_{cq})$ with respect to the tripartition $R\!:\!C\!:\!D$, defined as
\begin{equation}
\label{eq:def_j3}
\scalebox{0.9}{$\displaystyle
    \mathbf{j}_3^{\rm acc}(\tilde{\rho}) \equiv
    I_{\rm acc}(R:C)_{\tilde\rho}
    + I_{\rm acc}(R:D)_{\tilde\rho}
    - I_{\rm acc}(R:CD)_{\tilde\rho}\,,
$}\end{equation}
\addLI{where each summand in the previous equation has the same form as in \eqref{eq:notation_Iacc}.}
Note how this definition automatically implies a common input encoding for all three accessible informations, via the choice of $\rho_{cq}$, and thus $\tilde\rho$.
Finally, the accessible tripartite information is to be defined maximizing $\mathbf{j}_3^{\rm acc}(\tilde\rho)$ with respect to all possible input classical-quantum states $\rho_{cq}$ --- that is, more explicitly, with respect to all possible choices of weights $p_0,p_1$ and states $\rho_0,\rho_1$:
\begin{equation}\label{eq:I3acc_joint_optim}
    J_3^{\rm acc}(R:C:D) \equiv
    \sup_{\rho_{cq} } \mathbf{j}_3^{\rm acc}((I\otimes\Phi)(\rho_{cq})).
\end{equation}
\begin{figure}[h]
    \centering
    \includegraphics[width=\linewidth]{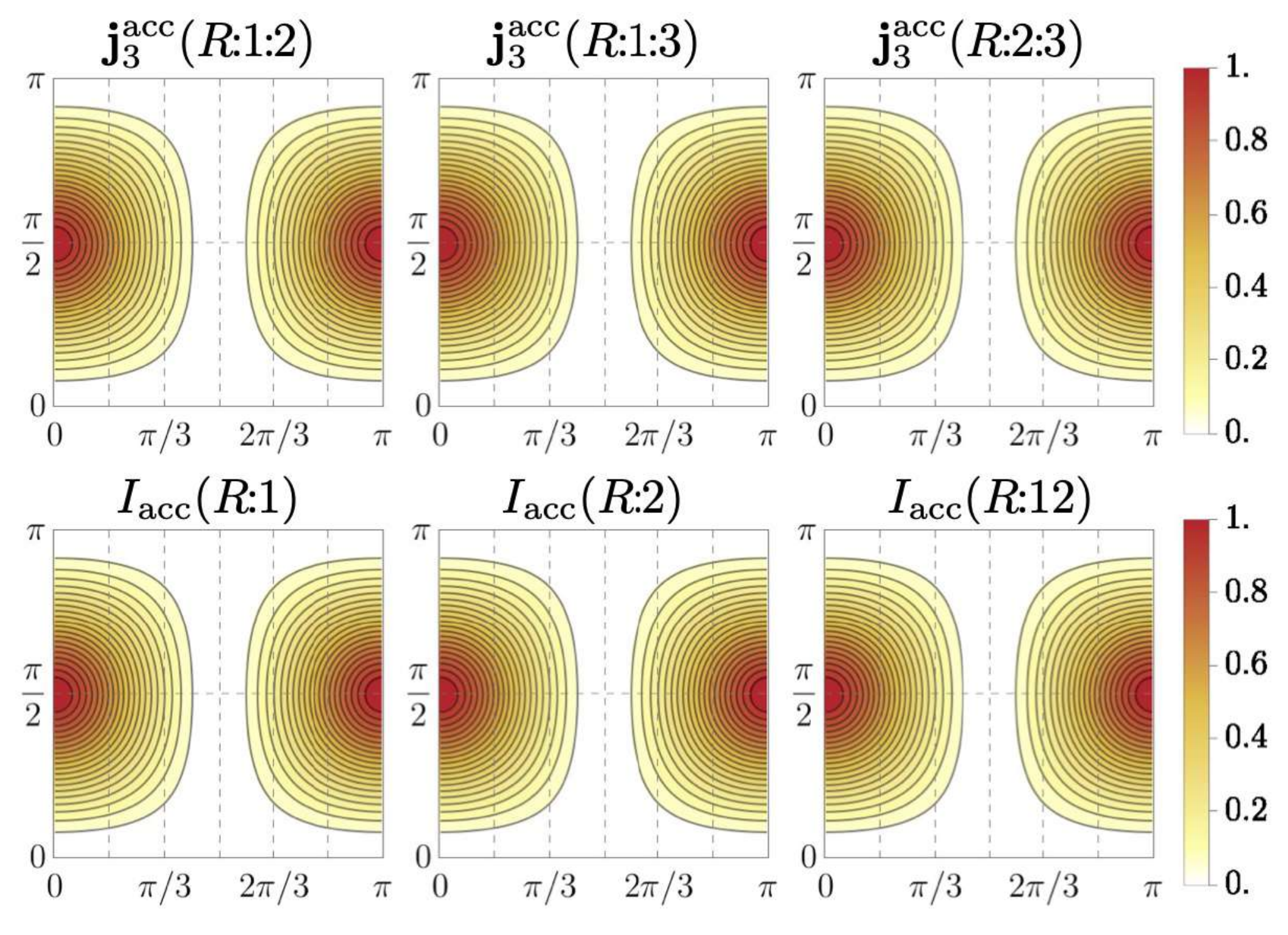}
    \caption{\Cref{eq:def_j3} evaluated as a function of input basis for the isometry defined in~\cref{eq:example_I3_pos}. First row reports the three different pair considered in the output (i.e. $\{1,2\},\{1,3\},\{2,3\}$). In the second row the three addenda $\small
    I_{\rm acc}(R:C)_\rho,\,
    I_{\rm acc}(R:D)_\rho,\,
    I_{\rm acc}(R:CD)_\rho$ relative to the first pair $\{1,2\}$}
    \label{fig:accessible_tripartite_infos_I3pos}
\end{figure}
% While performing the optimization underlying this quantity is nontrivial,
For ease of presentation, we consider here the results obtained assuming an encoding with balanced input distributions, \textit{i.e.}~$p_0{=}p_1{=}1/2$.
We then find, as also previously calculated explicitly, that first and second output qubits are optimally correlated with different input bases.

\begin{figure}[tbh]
    \centering
    \includegraphics[width=\linewidth]{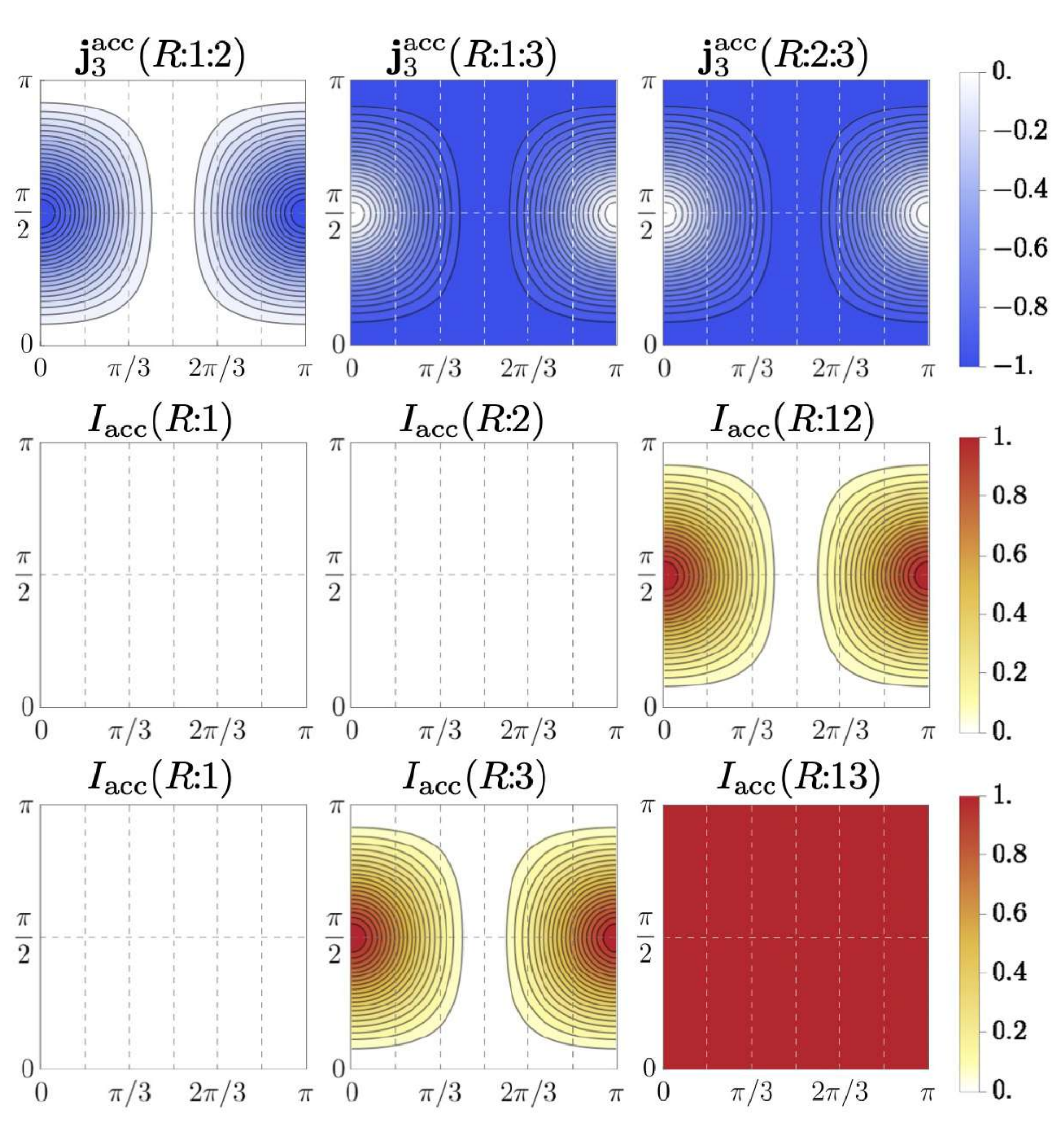}
    \caption{\Cref{eq:def_j3} evaluated as a function of input basis for the isometry defined in~\cref{eq:example_I3_neg}. First row reports the three different pair considered in the output (i.e. $\{1,2\},\{1,3\},\{2,3\}$). In the second and third rows the three addenda $\small
    I_{\rm acc}(R:C)_\rho ,
    I_{\rm acc}(R:D)_\rho ,
    I_{\rm acc}(R:CD)_\rho$ relative to the pairs $\{1,2\}$ and $\{1,3\}$ respectively.}
    \label{fig:accessible_tripartite_infos_I3neg}
\end{figure}
\addLI{This is shown in~\cref{fig:accessible_tripartite_infos_W3,fig:accessible_tripartite_infos_I3pos,fig:accessible_tripartite_infos_I3neg}, where in each plot we give the value of accessible and encoding-dependent tripartite information for the various considered examples, highlighting the dependence of these quantity on the choice of basis used to encoding information in the input.}
\addLI{Each point in the three plots corresponds to a choice of input encoding basis, with the possible orthonormal bases parametrized via the standard spherical coordinates in the Bloch sphere: horizontal axis corresponds to the azimuthal angle $\varphi\in[0,\pi]$, and vertical axis to $\theta\in[0,\pi]$. We can restrict to considering the range $[0,\pi]$ for $\varphi$ because angles in $[\pi,2\pi]$ correspond to identical encoding bases.}
In particular, the enconding-dependent accessible information is roughly $J_3^{\rm acc}(R:1:2)\approx0.186$, this optimal value being reached in correspondence of the basis $(\theta=\pi/2,\phi=5\pi/6)$,
%In particular, we find that using the encoding basis corresponding to $\phi=\pi/3$, the accessible tripartite information is roughly $J_3^{\rm acc}(R:1:2)\approx0.18$,
% \cLI{also, do we need a name for this ``fixed'' accessible tripartite information?}
in stark contrast with the corresponding value of $I_3^{\rm acc}(R:1:2)\simeq -0.06$.
This result is easily understood as a consequence of the fact that, in computing $I_3^{\rm acc}$, one does not take into account that the input encodings optimizing correlations with respect to different output qubits might not be compatible with one another, and we thus end up overestimating the amonut of local correlations.

%\addRev{Add a recap of the two results plus I3. Comment why the three results are different (I3 is affected by discord+basis choice problems, I3acc only by basis choice).
%$I_3=-\frac{3}{2}\log_2(4/3)\approx-0.62$}

\Cref{eq:I3acc_joint_optim} thus gives a way to more accurately assess the amount of information encoded in some fixed input basis that is hidden in nonlocal correlations between different output subsystems.
As we showed for the isometry in~\cref{eq:W3_isometry}, taking care of using the same encoding basis when computing the tripartite information can in fact produce results that are rather different than those obtained using more standard approaches, and that arguably more accurately reflect the scrambling properties of the evolution.
This is an aspect of QIS that has not currently been appreciated in the literature, but can significantly affect estimates of QIS.

\addLI{The richness of the phenomenon of quantum information scrambling can be also appreciated looking at \cref{table:summary_of_eqs}, summarizing the values of the various definition of tripartite information for all the examples studied. Even at a cursory look, it is evident that the accessible tripartite informations (both $I_3^{\rm acc}$ and $J_3^{\rm acc}$) are more suitable to capture the complexity of the QIS properties of different dynamics.
% return a more diversified scenario with respect to the one depicted using $I_3$. 
For instance, using accessible tripartite informations highlights how in many cases the way information is scrambled depends on the output tripartition.
% of the information that can be actually retrieved (for isometries \cref{eq:example_I3_van} and \cref{eq:ex_I3_-1}).
}

\begin{figure}[tbh]
    \centering
    \includegraphics[width=\linewidth]{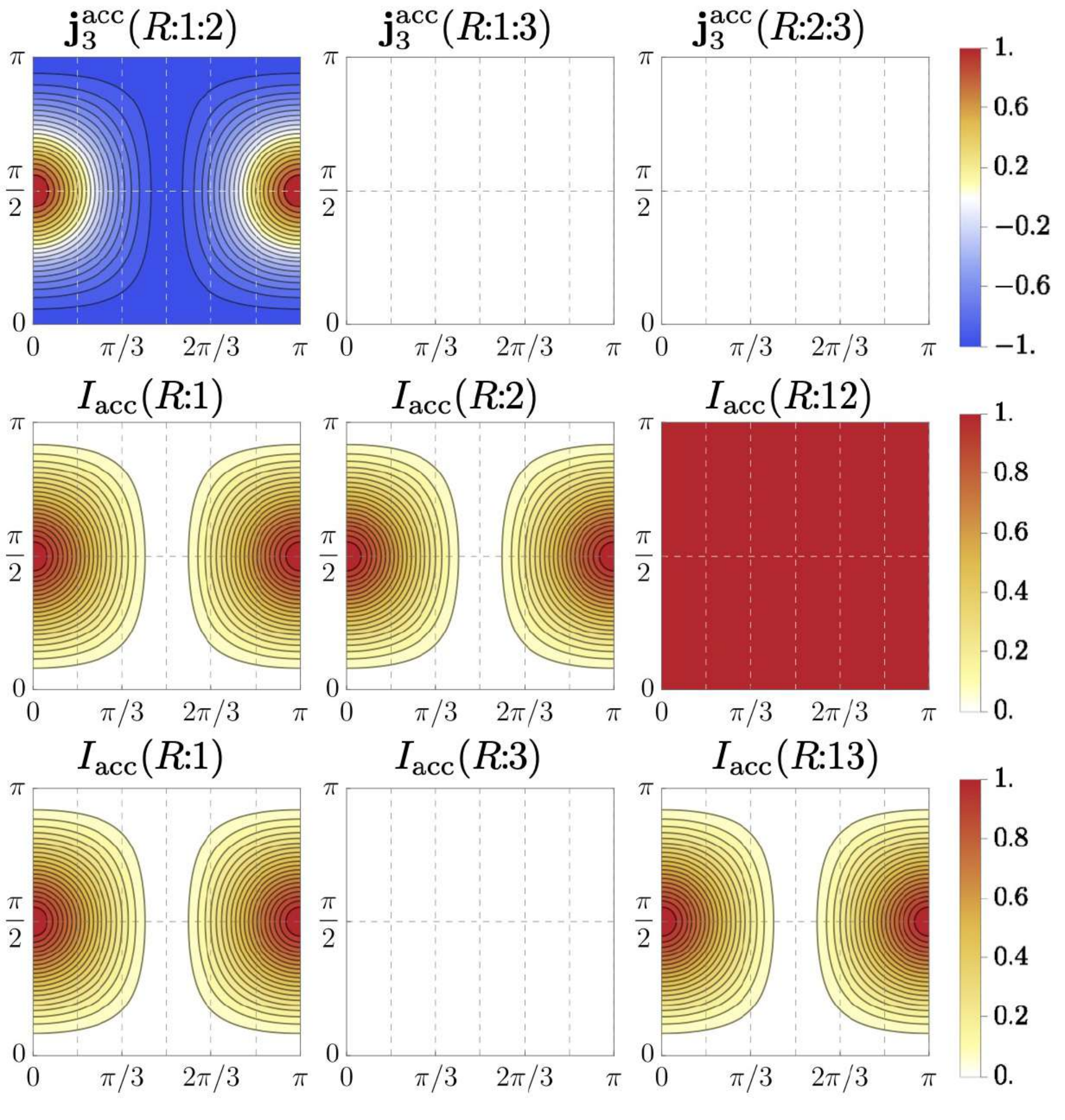}
    \caption{\Cref{eq:def_j3} evaluated as a function of input basis for the isometry defined in~\cref{eq:example_I3_van}. First row reports the three different pair considered in the output (i.e. $\{1,2\},\{1,3\},\{2,3\}$). In the second and third rows the three addenda $\small
    I_{\rm acc}(R:C)_\rho ,
    I_{\rm acc}(R:D)_\rho ,
    I_{\rm acc}(R:CD)_\rho$ relative to the pairs $\{1,2\}$ and $\{1,3\}$ respectively.}
    \label{fig:accessible_tripartite_infos_I3van}
\end{figure}

\section{Conclusions}
\label{sec:conclusions}

% \parTitle{Issues intrinsic to using QMIs to quantify QIS}
We showed, providing several explicit examples, that the merits of using the tripartite information of the Choi-like state $\rho_U$ to quantify the scrambling properties of the associated unitary evolution $U$ are questionable.
This approach does not allow to capture the complexity of the phenomenon of information scrambling, and often fails to accurately assess the amount of correlations spreading among output degrees of freedom, in good part due to the contributions of unaccessible correlations in quantum mutual informations.

\addLI{More specifically, the tripartite information is defined a sum of QMIs between a register $R$, carrying information about the input encoding, and various subsystems of the output, denoted in the main text as $C,D$ and $CD$. The negativity of  $I_3(R:C:D)$ is usually interpreted as the fact that we can retrieve more input information from $C$ and $D$ as a whole. In other words, part of the input information is stored into correlations between $C$ and $D$. However, being defined in terms of QMIs, the tripartite information does not
directly quantify correlations that are accessible via measurements on the output system.
% distinguish between {\it classical} and {\it nonclassical} correlations, possibly overestimating the amount of correlations actually accessible through local measurement.
Furthermore, the stripartite information is not sensitive to the choice of the input basis, while the amount of hidden information often depends crucially on the choice of encoding on input states, as we showed explicitly in several examples.}

% \parTitle{Fixing these issues via accessible quantities}
To avoid these shortcomings, we put forward the use of \textit{accessible} mutual informations, and a corresponding accessible tripartite information, to more precisely quantify accessible correlations between inputs and outputs.
We show that this is equivalent to the previous use of the quantum mutual information of the Choi states $\rho_U$, except for an application of a suitable fully dephasing channel before computing the tripartite information.
% \addRev{add explicit reference to the equations in the text}
We study several examples showing that this new definition provides results that more accurately reflect what we mean to label as QIS.
\addLI{More specifically, in section \ref{sec:accessible_I3}, we provided a number of examples where the amount of scrambling estimated by tripartite information is in stark contrast with the results obtained using the accessible triparite information. For instace, the isometry \eqref{eq:example_I3_van} has $I_3\!=\!0$ for any choice of (non-trivial) tripartition of the output. However, an analogous investigation using the accessible tripartite information reveals a more complex scenario: some output subsystems turn out to have $I_3^{\text{acc}}=-1$, meaning that the standard tripartite information was overestimating the information actually accessible. In other cases, as for the isometry \eqref{eq:ex_I3_-1} we stressed how the standard tripartie information is not able to capture the dependence on the choice of encoding basis of QIS. All those examples aim to point out how accessible tripartite information charts more faithfully the way information is stored and distributed among various subsystems of the output.}
\addLI{Although most of the presented examples use $2$-dimensional qubit states, the formalism works in general dimensions. We leave the exploration of possible phenomenology arising from the study of higher dimensions for future work.}

% \parTitle{Encoding in different bases}
Finally, in section \ref{sec:encoding_bases}, we showed that the use of accessible informations to quantify QIS reveals an additional interesting feature of how information can spread in quantum systems.
Namely, the fact that different output systems can be optimally correlated with the input states with respect to different bases.
This suggests yet another arguably more appropriate way to quantify QIS, by computing tripartite informations with respect to common input encodings.
This fascinating phenomenon suggests that further study is warranted to better understand the nature of QIS.

% \parTitle{Outlooks}
Our analysis paves the way for a novel approach to QIS. In part, it would be highly interesting to see to what degree results obtained with the standard approach using quantum mutual informations still hold when accessibility issues are taken into consideration.
Furthermore, given the known relations between OTOCs and tripartite information, the question arises as to whether accessibility issues are also present when OTOCs are used, and more generally whether a more direct operative interpretation of OTOCs is also possible.
Finally, the observed phenomenon of different output systems ``reading'' the input in different bases begs the question of what is the precise degree to which this is possible at all --- we know that some limitations must exist due to Holevo's theorem: it cannot be possible, for example, for different output systems to be maximally correlated with the input in different bases, as that would allow to measure the state in multiple incompatible bases at the same time.
We leave all these additional aspects as venues for future work.
\acknowledgments

L.I. acknowledges support from MUR and AWS under project 
PON Ricerca e Innovazione 2014-2020, ``calcolo quantistico in dispositivi quantistici rumorosi nel regime di scala intermedia" (NISQ - Noisy, Intermediate-Scale Quantum). GLM and GMP acknowledge funding from the European Union - NextGenerationEU through the Italian Ministry of University and Research under PNRR - M4C2-I1.3 Project PE-00000019 "HEAL ITALIA" (CUP B73C22001250006 ) 
SL and GMP acknowledge support by MUR under PRIN Project No. 2017 SRN-BRK QUSHIP. DAC acknowledges support from the Horizon Europe EIC Pathfinder project QuCoM (Grant Agreement No.101046973)

\clearpage
\newpage

\bibliography{bibliography}
% \printbibliography

\end{document}